# Are the second law principles of Caratheodory and Kelvin equivalent?

P. Radhakrishnamurty

42, Shivakrupa 1st Floor, 3A Main, Manjunathanagar, 2nd Phase
Bangalore – 560010, India. e mail: padyala1941@yahoo.com

**Abstract**

Lord Kelvin's postulate is a standard form of statement of the second law of thermodynamics. Caratheodory's principle is also considered as a different form of this law. Landsberg deduced Caratheodory's principle from Kelvin's postulate and Dunning-Davies deduced its converse to show the equivalence of the two principles, and these deductions remained forerunners for later works. Here we show that these deductions are flawed and therefore, invalid. The fallacies in these works arose due to the fact that these authors had taken an inadmissible form of statement of Kevin's postulate and had considered an inappropriate cyclic process, for their deductions. The equivalence between Caratheodory's principle and Kelvin's principle of the second law, therefore, remains unproved.

______________________________________________________________________________

## Introduction

Until the beginning of the 20th century, the exposition of the laws of thermodynamics was considered as lacking mathematical rigor [1]. In 1909, Caratheodory, at the instance of the famous physicist, M Born, provided that rigor and succeeded in extracting the quintessential physical basis for the second law of thermodynamics [2]. He set out, by defining the thermal concepts (heat, temperature etc.) in mechanical terms (mass, force, pressure, etc.). Thus the terms heat and temperature are not used in his statement of the laws of thermodynamics. However, his analysis was considered difficult to understand. Later, Born explained Caratheodory's analysis in clearer terms [3].

The first law statement in Caratheodory's formulation runs thus [2]: With every phase $\varphi_i$ of a system S it is possible to associate a function $\varepsilon_i$ of the quantities $V_i$, $p_i$, $m_{\kappa i}$ (where the subscript $i$ denotes the phase and $\kappa$ the component, $v$, $p$, $m$ denote volume, pressure and mass respectively), which is proportional to the total volume $V_i$ of the phase and which is called its internal energy. The energy of the system is the sum of the energies of the phases. Thus, $\varepsilon = \Sigma \, \varepsilon_i$.

Alternately [3], The work necessary to transfer a closed system from one equilibrium state to another equilibrium state by an adiabatic process depends only on these two states, not on the process of application of work (this statement is based on Joule's experimental results on the mechanical equivalent of heat). During an adiabatic process, the sum of the work, $W$ and the energy difference vanishes. If $\underline{U_f}$ and $U_i$ are the final and initial values of $U$, $(U_f - U_i) + W = 0$. ($U$ is used as a standard symbol for the internal energy, in place of $\varepsilon$, used by Caratheodory).

The second law statement in Caratheodory's formulation runs thus [2]: In every arbitrarily close neighborhood of a given state there exist states that cannot be approached arbitrarily closely by adiabatic processes.

# Deduction of Caratheodory's principle from Kelvin's postulate

In order to confirm a new formulation of the second law to be a valid one, it is customary to demonstrate that it is deducible from either Kelvin's statement or Clausius statement of the second law – the two of which are known to be deducible from one another.

During the mid nineteen sixties Landsberg [4] deduced Caratheodory's principle from Kelvin's principle (postulate). Later, Dunning-Davies [5] attempted to show the equivalence of Caratheodory's principle and Kelvin's principle (postulate) of the second law by deducing one principle from the other. A perusal of recent literature [6-8] shows, later works also treated the issue of equivalence of the two principles on similar lines. As a result, Caratheodory's principle and Kelvin's principle are considered today as equally valid and alternate forms of statement of the second law.

We do not find in the literature, to the best of our knowledge, reports where these claims have been contested. Caratheodory's principle of the second law remains a standard form of statement of the second law similar to those of Kelvin and Clausius.

We demonstrate below that the deductions of Landsberg and Dunning-Davies are flawed and are therefore invalid. This, we show by demonstrating that: 1. The statement for Kelvin's principle they considered in their deductions was an inappropriate one, and 2. The cyclic process they considered in their deductions was also an inappropriate one; one, which in no way corresponds to the cyclic process on which Kelvin's principle is based. As such the deductions failed to deduce the equivalence between the principles of Caratheodory and Kelvin. Thus, the equivalence between Caratheodory's principle and Kelvin's principle of the second law remains unproved.

# Landsberg's deduction of Caratheodory's principle from Kelvin's principle

Let us first consider Landsberg's work [4], wherein he claimed to have deduced Caratheodory's principle from Kelvin's principle for the first time. (Later works on this issue take this publication as the land mark).

To deduce Caratheodory's principle from Kelvin's principle, Landsberg took an inappropriate statement for Kelvin's principle, quoted below:

"It is impossible to convert an amount of heat completely into work in a cyclic process without at the same time producing other changes."

This, however, is not Kelvin's principle. The standard form of Kelvin's principle is one such as given by Fermi [9], quoted below:

"A transformation whose only final result is to transform into work heat extracted from a source which is at the same temperature throughout is impossible. (Postulate of Lord Kelvin)".

The vital part of Kelvin's postulate, is the insistence on a heat source which is at the same temperature throughout (a heat reservoir, for example). Kelvin's principle involves one-temperature cyclic processes (1-T cycles). It does not apply to cyclic processes that do not at all involve heat interaction (0-T cycles), and to cyclic processes that do involve heat interaction but at more than one temperature (n-T cycles, n >1).

The absence of any mention of temperature(s) at which heat interaction occurs (if it occurs at all) in Landsberg's statement of Kelvin's principle makes it an inappropriate statement of Kelvin's principle. Therefore, a process that violates such a statement cannot be construed as violating Kelvin's postulate.

## Landsberg's cycle process and its analysis

Landsberg considered a thermodynamic phase space the points in which represent equilibrium states of a system. Then he supposed that it contained a point A possessing a neighbourhood N, all points of which could be reached adiabatically from A. The coordinates of the phase space were the internal energy $U$ and deformation coordinates $v_1$, $v_2$, ...... .

We now look at Landsberg's cyclic process: Keeping $v_1$, $v_2$, ...... fixed, the system was taken from a point B ($U_B < U_A$) in N to the point A. With the $v_i$ constant, no work was done. The increase in $U$ was due to heat $Q$ ($> 0$) having been supplied[*]. Because B lies in N, the system can be returned from A to B by an adiabatic process, using appropriate changes of deformation coordinates. The drop in $U$ is now entirely due to mechanical work which has been done by the system. That completes a cyclic process, B A B.

Application of the first law to the cycle B A B demands that $W = Q$, so that heat $Q$ has been completely converted into work. Landsberg argued that such a result contradicted Kelvin's principle and therefore, was an impossible result.

According to Landsberg, the impossibility arose due to the assumption that all points in the phase space that lie in the neighbourhood of an arbitrarily chosen point A are reachable from A by an adiabatic process. Therefore, he concluded that, points such as A cannot exist. Hence his deduction: "in every neighbourhood of every point C in thermodynamic phase space there are points adiabatically inaccessible from C."

## Landsberg's cycle is an inappropriate form of Kelvin's cycle

The process of complete conversion of heat into work in a cyclic process without at the same time producing other changes, contradicts Kelvin's principle if, and only if, heat interaction does occur with surroundings, and does so, at a single temperature during the cyclic process. So, if Landsberg's cycle involved heat interaction between the system and surroundings, then $T_A$ must be equal to $T_B$ so that his cycle corresponds to Kelvin's cycle. This is not the case however, with the cycle B A B considered by Landsberg (since $T_A \neq T_B$). Hence, Kelvin's principle does not apply to such a cyclic process. Stated differently. Landsberg's cyclic process is not an impossible cyclic process according to Kelvin's principle.

---

[*] Two possibilities arise here. 1. Heat *Q* is supplied through heat interaction with surroundings. In such a case, that heat interaction occurs not at one temperature but over a range of temperatures, since $T_A \neq T_B$. 2. Heat *Q* is supplied by a mechanical process (such as Joule's process), that is, by an adiabatic process. Therefore, the cycle B A B is either not a one-temperature cycle (as Kelvin's cycle should be) or it is an adiabatic cycle. In either case the process would not apply to Kelvin's principle, and Landsberg's deduction is deemed to be erroneous. Detailed analysis is given in the text.

Alternately, if there is no heat interaction with surroundings during the whole cyclic process, that is, if B A B is a 0-T cycle, then also Kelvin's principle does not apply to such (adiabatic) cycle.

Thus, since Landsberg's cyclic process does not correspond to the cyclic process of Kelvin's principle, his conclusion that the heat $Q$ got completely converted into work in his cyclic process does not violate Kelvin's principle. Hence, Landsberg's conclusion, that in every neighbourhood of every point C in thermodynamic phase space there are points adiabatically inaccessible from C, is erroneous. Therefore, Landsberg's deduction of Caratheodory's principle from Kelvin's principle is invalid.

Dunning-Davies [5] attempted to deduce Kelvin's principle from Caratheodory's principle to prove thereby, in conjunction with Landsberg's deduction, the equivalence of the two principles. He also took Kelvin's principle to be one similar to that taken by Landsberg (see below). He considered Landsberg's cycle, run in the reverse direction. As shown above, such cycles do not correspond to cycles in Kelvin's postulate. Dunning-Davies deduction that considers Landsberg's cycle in the reverse direction, fails to deduce Kelvin's principle from Caratheodory's principle. Therefore, equivalence of Caratheodory's principle and Kelvin's principle of the second law remains unproved.

To conclude, we reiterate that the essence of Kelvin's postulate lays in the fact, that it asserts the impossibility of operation of 1-T cyclic processes, and only 1-T cyclic processes, in a certain direction – the direction that transforms heat into work. 1-T cyclic processes are possible only in one direction but not in the opposite direction. In other words, 1-T cyclic processes are necessarily irreversible. Other cyclic processes such as 0-T cyclic processes are necessarily reversible, (that is, possible to run in both forward and backward directions) and n-T cycles, where, n >1, are not necessarily irreversible, (that is, not impossible to run in both forward and backward directions, in every case).

Several other publications [6-8] also adapted similar statements for Kelvin's principle of the second law. Surprisingly, even Born, adopted a similar statement! Consequently, their results are deemed to be invalid. We quote their versions of Kelvin's statements below.

"….. Kelvin's principle that it is impossible to transform an amount of heat completely into work in a cyclic process in the absence of other effects." [5].

Kelvin's formulation of the second law, that *"no cycle can exist whose net effect is a total conversion of heat into work."* [6].

"Kelvin (and Planck): No process is possible, the sole result of which is that a body is cooled and work is done." [7, 8]

## Acknowledgement

I wish to express my sincere thanks to Prof. Jeremy Dunning-Davies who, on my request, kindly sent me his works connected with thermodynamics and the paper of Prof. Landsberg; and also for the very useful email correspondence we had on the deduction of Caratheodory's principle from Kelvin's principle and the equivalence of the two principles. It must be mentioned though, our views on the issue diverged.